\documentclass[twocolumn,showpacs,preprintnumbers,amsmath,amssymb,
groupedaddress]{revtex4}

\usepackage{graphicx}
\usepackage{dcolumn}
\usepackage{bm}

\begin{document}

\title{Route from spontaneous decay to complex multimode dynamics in cavity QED}

\author{Dmitry O.~Krimer}
\email{dmitry.krimer@gmail.com}
\affiliation{Institute for Theoretical Physics, Vienna University of Technology, A-1040, Vienna, Austria, EU}
\author{Matthias Liertzer}
\affiliation{Institute for Theoretical Physics, Vienna University of Technology, A-1040, Vienna, Austria, EU}
\author{Stefan Rotter}
\affiliation{Institute for Theoretical Physics, Vienna University of Technology, A-1040, Vienna, Austria, EU}
\author{Hakan E.~T\"ureci}
\affiliation{Department of Electrical Engineering, Princeton University, Princeton,  New Jersey 08544, USA}

\begin{abstract}
We study the non-Markovian quantum dynamics of an emitter inside an open
multimode cavity, focusing on the case where the emitter is resonant with 
high-frequency cavity modes. Based on a Green's function technique suited for
open photonic structures, we study the crossovers between three distinct regimes
as the coupling strength is gradually increased: (i) overdamped decay with a time scale
given by the Purcell modified decay rate, (ii) underdamped oscillations with a time scale
given by the effective vacuum Rabi frequency, and (iii) pulsed revivals. The final
multimode strong coupling regime (iii) gives rise to quantum revivals of the atomic
inversion on a time scale associated with the cavity round-trip time.  
We show that the crucial parameter to capture the crossovers between these regimes
is the nonlinear Lamb shift, accounted for exactly in our formalism.
\end{abstract}


\maketitle

\section{Introduction}
Controlling the emission properties of quantum systems is at the heart of a number of fields ranging from quantum information processing to single-molecule spectroscopy. In solid-state cavity QED a substantial amount of experimental effort aims at designing highly structured photonic environments in the vicinity of the emitter to achieve a high level of control over its quantum dynamics \cite{KhitrovaGKKS2006, NodaFA2007, WallraffSBFHMKGS04, Agio2011}. Much of the earlier work focuses on the resonant coupling to a single confined mode of the photonic structure that has favorable emission properties, while coupling to the rest of the modes of the photonic environment is regarded as a parasitic influence and is either discarded or bulked into a total background spontaneous emission rate in the spirit of Ref.~\cite{CarmichaelBRKR89}. Recent trends in experimental work, however, point towards spatially highly complex and open photonic structures, where the delineation between a cavity and the radiative environment becomes highly blurred (see, e.g., \cite{Sapienza:2010vp, Ruijgrok2010, ChenAS2012}). Such situations are more effectively described through the local density of photonic states (LDOPS) \cite{Dung2000,Vats2002, WubsSL2004, OchiaiIS2006, Rao2008kj, Pierrat:2010ng, ChenSA2013}. This more powerful and potent theoretical approach has meanwhile fueled a great deal of research on light-matter interaction in fields ranging from cavity QED to photovoltaics \cite{Polman2012}, giving rise to what may be referred to as ``LDOPS engineering".

While recent theoretical works have recognized the potential of this method \cite{ WubsSL2004, OchiaiIS2006,Rao2008kj,Kristensen:2011pu} including those dealing with dispersing and absorbing media \cite{Dung2000,Khanbekyan:2008ap}, the lack of a suitable method which allows tackling the often complex non-Markovian dynamics of a two-level-like emitter in a leaky photonic structure was a significant hurdle in revealing novel phenomena that may be at play in a host of modern-day light-confining structures such as periodic \cite{NodaFA2007}, deterministic aperiodic \cite{Boriskina:2009eq}, and disordered photonic media \cite{Sapienza:2010vp}, as well as nanoplasmonic systems \cite{Chang:2006jq}. Here we present a formalism for computing the full quantum dynamics of emitters in arbitrarily complex photonic structures based on a single Volterra equation with a spectral function proportional to the LDOPS. We then illustrate the possibility of calculating the LDOPS of open and complex photonic structures employing the non-Hermitian set of Constant Flux states (CF states) that have been introduced in Ref.~\cite{Tureci06} to describe steady-state lasing characteristics of lasers.  Based on this powerful tool we explore the dynamics of a quantum emitter in the multimode regime, i.e., when the emitter couples to several modes of the cavity. This regime is notoriously difficult because it leads to highly complex non-Markovian dynamics, but it best illustrates the potency of the method outlined here to provide insight into the various possible time scales of the emitter dynamics. In particular, we discuss a series of cross-overs between three dynamical regimes as the coupling strength of the emitter is increased. Some of the aspects of these regimes have been discussed before in the literature within the limited scope of a variety of methods \cite{WW1930,Purcell,Jaynes_Cumm,Milonni,Plank,Dung2000,Carmele013}. The beauty of our approach which we present here, is that it provides a unified description, a thorough understanding and a classification for all of these regimes, with a key parameter being the nonlinear Lamb shift.

\begin{figure}[!t]
\includegraphics[angle=0,width=0.85\columnwidth,clip=true]{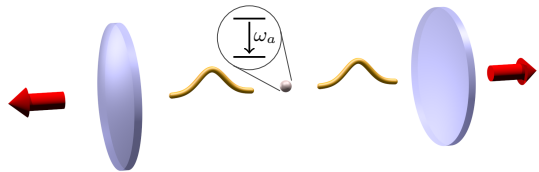}
\caption{(color online) Two-level system (TLS) with transition frequency $\omega_a$ inside an open cavity.}
\label{fig_contour}
\vspace*{0.cm}
\end{figure}
\section{Theoretical Model}
\label{Sec_Theor_Model}
The system we study is a typical cavity QED setup consisting of a two-level system (TLS) with transition frequency $\omega_a$ placed inside a cavity. The method we present here is valid for an arbitrarily complex open cavity geometry but for the sake of transparency we discuss here a Fabry-P\'{e}rot cavity formed by two highly reflecting mirrors, see Fig.~\ref{fig_contour}. To describe the excitation dynamics of the TLS we start with the familiar Hamiltonian written in terms of the modes-of-the-universe approach \cite{Glaub}, which makes no distinction between the cavity and its environment, ${\cal H}=(\hbar \omega_a/2) \cdot\sigma_z+\int d\omega  \hbar \omega a^{\dag}(\omega)\, a(\omega)+\hbar \sqrt{\gamma/\pi}\cdot\int d\omega [g(\omega, {\bf r})\, a(\omega)\sigma^+\!+\!g^{\star}(\omega,{\bf r})\, a^{\dag}(\omega)\sigma^-]$. Here 
$a^{\dag}(\omega)$ and $a(\omega)$ are standard creation and annihilation 
operators of a photon and $\sigma^+,\,\sigma^-,\,\sigma_z$ 
are the Pauli operators associated with the TLS. The interaction part of ${\cal H}$ is written in the electric dipole and rotating-wave approximation, where $g(\omega, {\bf r})$ are the coupling amplitudes, and $\gamma$ stands for the coupling strength proportional to the dipole moment squared. Due to the rotating 
wave approximation, non-resonant terms ($\propto a_\lambda\sigma^-,\,a_\lambda^\dag \sigma^+$) are absent in this Hamiltonian, such that the number of excitations is
conserved. We can thus make the following ansatz for the time evolution 
of the system, $|\Psi(t)\rangle=c(t)e^{-i \omega_a t/2} |u\rangle |0\rangle +\int d\omega c_{\omega}(t) |l\rangle |1_{\omega}\rangle e^{-i(\omega-\omega_a/2) t}$, 
where the ket-vectors $|u\rangle$ and $|l\rangle$ stand for
the atom in the upper and lower states, respectively, and the ket-vectors $|0\rangle$
and $|1_{\omega}\rangle$ represent the vacuum state and a
single photon with the frequency $\omega$. Solving the Schr\"odinger equation with this ansatz, ${\cal H}|\Psi(t)\rangle=i\hbar\partial_t|\Psi(t)\rangle$, we arrive at the following Volterra 
equation for the excited state amplitude of the TLS, $c(t)$,
\begin{equation}
\label{Volterra_eq}
\dot c(t)=-\dfrac{\gamma}{\pi}\int_0^t dt' \int_0^\infty d\omega F(\omega)
e^{-i (\omega-\omega_a)(t-t')}c(t'),
\end{equation}
where $F(\omega)=\rho({\bf r}_a,\omega)\cdot |  g(\omega)|^2$ is the spectral function, featuring the local density of photonic states (LDOPS), $\rho({\bf r}_a,\omega)$, evaluated at the emitter position ${\bf r}={\bf r}_a$ and $g(\omega)$ is the frequency dependent coupling amplitude.

Note that Volterra equations as above have already been used (i) for describing a single discrete energy level coupled to a featureless continuum of states \cite{Cohen} as well as (ii) for the case of a TLS coupled to dispersing dielectrics \cite{Dung2000, Bondarev}. In the former case (i) a very intuitive graphical analysis was presented including, however, a spurious integral extension towards negative frequencies. In the second case (ii) the solutions were calculated explicitly without, in turn, the insight provided by the modes of the corresponding open cavity geometry. In the following we introduce a method that is general enough to overcome the limitations of both approaches. 
 
To make contact with the physics of an open cavity, we first evaluate the LDOPS for a one-dimensional cavity of length $L$ bounded at $x=0,L$ by two thin semi-transparent mirrors modelled by dielectric slabs of width $d\ll L$ with refractivity index $n$ (see Fig.~\ref{fig_contour}). In what follows we use units where the speed of light, $c = 1$. We also normalize $x$ to $L$, measure time $t$ in units of half the cavity round trip time, and frequency $\omega$ in units of it's inverse. In the limit of $n\rightarrow \infty$ and $d\rightarrow 0$ the mirror's transparency is characterized by a factor $\eta=n^2 d$ which is related to the frequency dependent mirror's reflection amplitude as $r(\omega)=i \omega \eta/(2- i \omega \eta)$ \cite{Viviescas_geom}. For such an open system the LDOPS is given exactly by the imaginary part of the Green's function \cite{economou}, $\rho(x_a,\omega)=-2 \omega\cdot  {\rm Im}\, G^+(x_a,x_a,\omega)/\pi$, where the retarded Green's function (labeled by $+$) satisfies the Helmholtz equation $\left(\partial_x^2+n^2 \omega^2\right)G^+(x,x_a,\omega)=-\delta(x-x_a)$ for all $x\in{\mathbb R}$. Note that, due to the openness of the cavity, the LDOPS is a continuous function, corresponding  to a continuum of extended modes which are notably different from the discrete set of cavity modes. An exact discrete spectral representation for the Green's function can however be obtained for the finite but open cavity geometry at the expense of introducing a non-Hermitian set of modes referred to as ``constant-flux'' (CF) states, recently introduced to laser physics \cite{Tureci06,TGRS08}. To compute the response to a monochromatic source at frequency $\omega$, CF states $\phi_m(x)$ have to be determined which satisfy $[\partial_x^2+n^2\omega_m(\omega)^2] \phi_m(x)  = 0$ with the outgoing boundary conditions $\partial_x \phi_m(x)= \pm i\omega \phi_m(x)$ at the right (with $+$) and left cavity boundary (with $-$). These states can be understood to carry a constant flux to infinity \cite{Tureci06}. The resulting non-Hermitian eigenvalue problem features complex eigenvalues $\omega_m$ and a complete set of right ($\phi_m$) and left ($\bar{\phi}_m$) eigenvectors which parametrically depend on $\omega$ and are biorthogonal to each other, $\int_0^L dx\, n^2 \bar{\phi}^*_m \phi_n=\delta_{mn}$. The spectral representation of the Green's function can then be constructed through $G^+(x,x',\omega)= -\sum_m \phi_m(x,\omega)\bar{\phi}^*_m(x',\omega)/[\omega^2-\omega^2_m(\omega)]$, resulting in a LDOPS in the middle of the cavity which consists of a series of peaks, one for each $m$.  In this picture it becomes intuitively clear that the peaks in the LDOPS, which the TLS  couples to, arise when (i) the frequency $\omega$ is close to one of the CF frequencies $\omega_m$ (see denominator in the Green's function) and (ii) when the CF eigenfunction $\phi_m$ has a sizeable value at the position $x_a$ of the TLS (see the numerator). The function  $g(\omega)$ which determines the coupling strength to the emitter is given by $|g(\omega)|^2= \pi/2  \cdot  \omega \,\,e^{-(\omega-\omega_a)^2/(2 \omega_c^2)}$, where we have introduced a gaussian cutoff at $\omega_c$. In our simulations we varied the cutoff frequency $\omega_c$ in a relative large frequency interval observing qualitatively similar behavior. In what follows we present results for $\omega_c=2 \omega_a$. Putting all terms together, the spectral function in our example is given by,
\begin{widetext}
\begin{equation}
\label{F_spectral}
F(\omega)=\dfrac{2 n^2 \omega  e^{-(\omega-\omega_a)^2}}{(n^2+1)^2-(n^2-1)^2\cos(2\omega
n d)+2 (n^4-1) \cos(\omega L)  \sin^2(\omega n d)+2n (n^2-1) \sin(\omega L)  \sin(2\omega n d)}.
\end{equation}
\end{widetext}
\section{Dynamical scenarios}
\label{Sec_Results}
We now proceed to solve Eq.~(\ref{Volterra_eq}) for a single excitation, initially stored in the TLS, $c(0)=1$. Applying a Laplace transform (see Appendix \ref{App_Lapl}), we derive the following expression for the amplitude $c(t)$,
\begin{equation}
\label{Eq_fin_c}
c(t)=\dfrac{\gamma}{\pi}\,e^{i\omega_at}\int_0^\infty d \omega\,U(\omega)\,
e^{-i \omega t},
\end{equation}
with the kernel function 
\begin{equation}
\label{Eq_Uk}
U(\omega)=\lim_{\varepsilon\rightarrow 0^{+}} 
\dfrac{F(\omega)}{\left[\omega-\omega_a-\gamma \delta( \omega)\right]^2+\left[\gamma
F(\omega)+\varepsilon\right]^2}
\,,
\end{equation}
and the nonlinear Lamb shift
\begin{equation}
\label{Lamb_shift}
\delta(\omega)=\dfrac{1}{\pi}\,\mathcal{P}\int\limits_0^{\infty}d\tilde\omega\,\dfrac{
F(\tilde\omega)}{\omega-\tilde\omega}\,,
\end{equation}
where $\mathcal{P}$ denotes the Cauchy principal value.
\begin{figure}
\includegraphics[angle=0,width=1.\columnwidth]{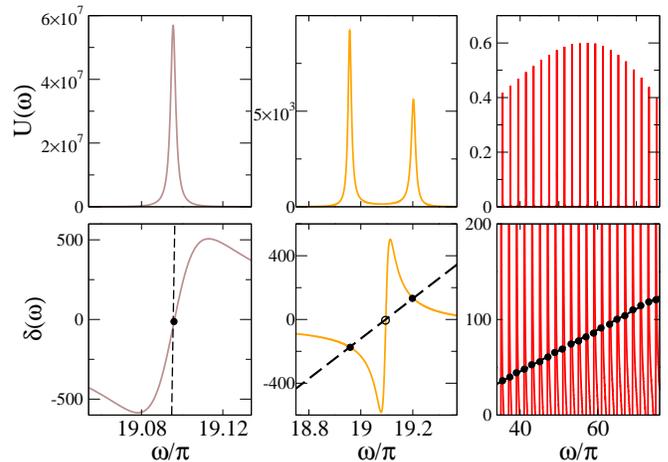}
\caption{(color online) 
Route from single- to multimode coupling regime for different coupling strengths
$\gamma$. {\it Upper row}: Dimensionless kernel function $U(\omega)$ [Eq.~(\ref{Eq_Uk})]. {\it Lower row}: Dimensionless nonlinear Lamb shift $\delta(\omega)$ [Eq.~(\ref{Lamb_shift})] for the same $\omega$-interval as above (note the different zooms for the three columns). 
{\it Left column}: weak coupling regime for $\gamma=4\cdot 10^{-6}$ with a single peak in $U(\omega)$
(Purcell modified spontaneous decay). {\it Middle column}: strong coupling regime for $\gamma=2.5 \cdot 10^{-3}$ with a well-resolved Rabi splitting in $U(\omega)$ (regime of damped Rabi oscillations).  {\it Right column}: Multimode strong coupling regime for $\gamma=1.44$ with a multi-peak structure in $U(\omega)$ consisting of almost equidistant peaks (regime of revivals). 
Filled circles label resonance values $\omega_r$ of the kernel
$U(\omega)$ occurring at the intersections between the Lamb shift
$\delta(\omega)$ and the dashed line $(\omega-\omega_a)/\gamma$. At empty circles (not shown in right column)
such intersections are non-resonant and do not lead to a corresponding peak in $U(\omega)$ (see text).
The transition frequency $\omega_a \approx 19 \pi$ of the TLS coincides with the 10th resonance of the spectral function $F(\omega)$ [Eq.(\ref{F_spectral})]. The reflectivity parameter $\eta = 0.1$ is such that the mirror reflectivity $|r(\omega_a)|^2=0.9$. Frequency $\omega$ is measured here in units of the inverse half the cavity round trip time.} 
\label{fig2_spectra}
\vspace*{0.15cm}
\end{figure}
\begin{figure}
\includegraphics[angle=0,width=1.\columnwidth,clip=true]{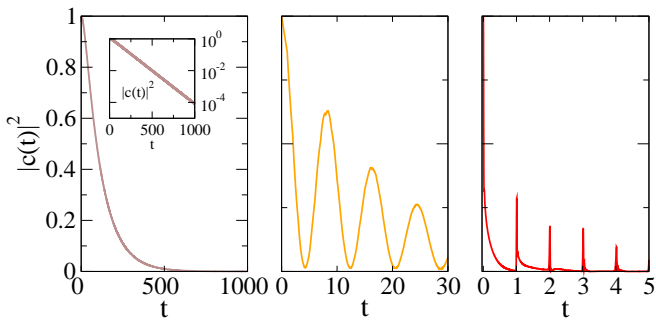}
\caption{(color online) Temporal evolution of the excited state probability $|c(t)|^2$ of the TLS 
for the three cases shown in Fig.~\ref{fig2_spectra}. Time $t$ is measured here in units of
half the cavity round trip time. {\it Left panel}: Weak coupling regime ($\gamma=4\cdot 10^{-6}$) featuring spontaneous decay 
(also shown in log-lin scale in the inset). 
{\it Middle panel}: Strong coupling regime ($\gamma=2.5 \cdot 10^{-3}$) with damped Rabi oscillations. {\it Right panel}: 
Multi-mode strong coupling regime ($\gamma=1.44$) featuring pulsed revivals at multiple integers of half the cavity round trip time.}
\label{fig3_evolution}
\vspace*{0.0cm}
\end{figure}
%
The dominant frequency components entering the dynamics of $c(t)$ are 
those which are resonant in the kernel function $U(\omega)$. A {\it
necessary} condition for such resonances to occur is that the first term 
in the denominator of $U(\omega)$ vanishes,
\begin{eqnarray}
\label{cond_res}
\dfrac{\omega_r-\omega_a}{\gamma}= \delta( \omega_r)\,.
\end{eqnarray}
This resonance condition is satisfied at the frequencies $\omega_r$, determined by the
intersection of the nonlinear Lamb shift $\delta(\omega)$ and a straight line $(\omega-\omega_a)/\gamma$  
(see a corresponding graphical analysis in \cite{Cohen} for a simple form of a continuum).
Since, according to Eq.~(\ref{Lamb_shift}), every resonance in $F(\omega)$ produces a 
dip followed by a peak in the Lamb shift, there may be several such intersections, 
corresponding to multiple solutions of Eq.~(\ref{cond_res}). The corresponding resonances in the kernel $U(\omega)$ can, however, be suppressed, whenever 
the spectral function $F(\omega)$ has a maximum at the same resonance frequency. This is the case if  the kernel $U(\omega)=1/[\gamma^2 F(\omega)]$ goes through a minimum at $\omega=\omega_r$.  

Based on these observations, we will now investigate the crossover from weak 
to strong coupling upon variation of the coupling strength $\gamma$;
all other parameters, like the spectral function $F(\omega)$ and the mirror's reflectivity factor $\eta$ will be left unchanged. At very weak coupling, $\gamma=10^{-4}$ (left panel of Fig.~\ref{fig2_spectra}), the straight line in Eq.~(\ref{cond_res}) is very steep and thus leads just to a single intersection, corresponding to a single resonance at $\omega_r \approx \omega_a$. All quantities in Eq.~(\ref{Eq_Uk}) can thus be evaluated at $\omega_a$ to very good accuracy and the kernel function reduces to a  Lorentzian centered around the slightly shifted frequency $\omega_a+\gamma \delta(\omega_a)$ with the width $\gamma F(\omega_a)$. By extending the integration limit in
Eq.~(\ref{Eq_fin_c}) to $-\infty$, we reproduce the Purcell modified exponential decay of the TLS inversion \cite{Purcell}, in good agreement  with a numerical solution of the Volterra equation, Eq.~(\ref{Volterra_eq}) (left panel in Fig.~\ref{fig3_evolution}). This is the overdamped dynamics of the TLS in the weak coupling limit of Cavity QED.

As $\gamma$ increases to $\gamma=2.5 \cdot 10^{-3}$ we enter the strong coupling regime, as indicated by the straight line now being flat enough to 
intersect the nonlinear Lamb shift at three points (middle panel of
Fig.~\ref{fig2_spectra}). Note that  these {\it three} intersections give rise to only {\it two} resonances $\omega_r$ in the kernel $U(\omega)$ since the middle frequency is very close to the resonance of $F(\omega)$ (see discussion above). As a
consequence, the kernel function $U(\omega)$ has a double peak structure that is characteristic of the single-mode vacuum Rabi splitting \cite{Jaynes_Cumm}. 
This energy splitting introduces a new frequency scale, the Rabi frequency, 
which is easily estimated from the resonance condition (\ref{cond_res}) to be
$\sqrt{2 \omega_a \gamma}$\,. The inverse of the
peak width provides the time scale at which the Rabi oscillations decay, as confirmed
by independent numerical solutions of Eq.~(\ref{Volterra_eq})
(middle panel of Fig.~\ref{fig3_evolution}).

\begin{figure}[!t]
\includegraphics[angle=0,width=1.\columnwidth,clip=true]{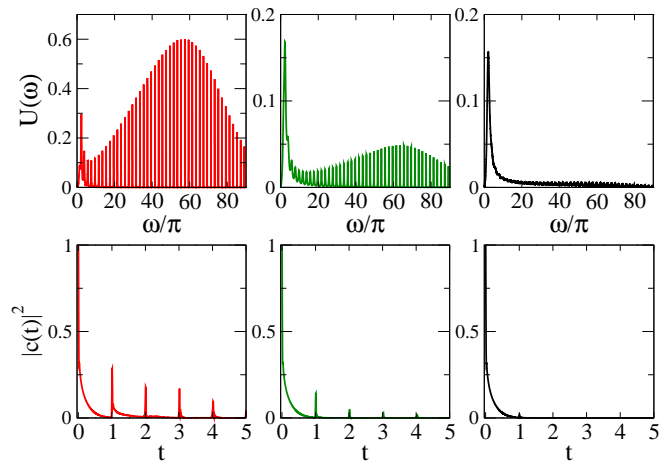}
\caption{(color online) Destruction of the multimode strong coupling regime by broadening of the
peaks in the spectral function, Eq.~(\ref{F_spectral}). {\it Left column}: Mirror reflectivity parameter
$\eta=0.9$ (as in right panel of Fig.~\ref{fig2_spectra}). {\it
Middle column}:  $\eta=0.3$. {\it Right column}: $\eta=0.015$. {\it Upper
row}: Dimensionless kernel function $U(\omega)$. {\it Lower row}: Corresponding excited state probability $|c(t)|^2$ of the TLS versus normalized time $t$. The transition frequency $\omega_a \approx 19\pi$ and the coupling strength $\gamma=1.44$ are the same as in the right panels of the previous two figures.}
\label{fig5_broadening}
\vspace*{0cm}
\end{figure}
With further increase of the coupling strength to $\gamma=1.44$, the straight line starts to
intersect neighboring resonances of $\delta(\omega)$, involving an increasing
number of cavity modes. Thus, within the multimode strong coupling regime it is
possible to couple to many cavity modes, including those that reside far away 
from the transition frequency $\omega_a$ (right panel of Fig.~\ref{fig2_spectra}).
Note that, similar to the situation above, only every second intersection 
with the Lamb shift produces a resonance in the kernel $U(\omega)$ which,
correspondingly, takes on a multi-peaked profile. If, as in our case, these
peaks also have an equidistant spacing to each other, 
then the interference between these resonant modes produces a train of pulses
in the probability of the excited state $|c(t)|^2$, corresponding
to pulsed revivals of the TLS inversion (right panel of
Fig.~\ref{fig3_evolution}). With the revival time being equal to half the cavity 
round trip time, the straightforward explanation of this phenomenon
is the repetitive emission and subsequent reabsorption of radiation by the TLS, when it
is back-reflected by the cavity boundaries. As such, this effect relies on the fact
that the phases acquired from all possible paths starting from and returning to 
the position of the TLS differ only by integer multiples of $2\pi$, a condition
which strongly depends on the position of the TLS in the cavity. Indeed, if we 
move the TLS away from the cavity center, a much more irregular type of dynamics
emerges (not shown). We also checked explicitly on the limitations that the
revival effect is subject to in terms of the cavity opening.  
For that purpose we performed numerical simulations for cavities with smaller 
values of the mirror's reflectivity factor $\eta=0.3,\,0.015$.  We observe that for decreasing values of $\eta$ the overlap between neighboring resonant peaks in $U(\omega)$ increases until they merge into a single wide resonance. As a result, the revivals in the inversion of the TLS die out when all resonances merge to a single peak, at which point  the decay will be just a simple exponential decay, no matter how large
the coupling strength $\gamma$ is. 

\section{Comparison with a system-and-bath formalism}

To verify the validity of the above results, we recalculated the temporal decay in all of the three regimes from above using a recently developed system-and-bath approach \cite{VivPRA03}. Under the rotating-wave and Born  approximations this approach can, in principle, also be reduced to a single Volterra 
equation as in Eq.~(\ref{Volterra_eq}). We have, however, been able to go beyond the Born approximation by solving a coupled set  of Volterra equations for the TLS and damped cavity modes explicitly numerically. These equations very well illustrate how costly it becomes numerically to obtain the solutions for the temporal decay without the Laplace transform employed above and how little insight one gets into these solutions when they have to be explicitly integrated in time. The fact that we obtain very similar results (for all the scenarios obtained above) with this more complex approach confirms, in turn, the validity of the simple and insightful strategy presented in the previous Sections.

\subsection{Total Hamiltonian}
Our starting point is a Hamiltonian which includes altogether five contributions 
from  the resonator, the external region, the TLS as well as from the interaction 
of the resonator with the external region and of the TLS with the
resonator (see Eq.~(81) in \cite{VivPRA03}),
\begin{eqnarray}
\nonumber
{\cal H}&=&\sum_\lambda \hbar \omega_\lambda a_\lambda^\dag 
a_\lambda+\int d\omega\,  \hbar \omega b^\dag(\omega)b(\omega)+\dfrac{\hbar \omega_a}{2}\sigma_z+
\\
&+&\hbar \sum_\lambda \int d\omega \left[ {\cal W}_\lambda (\omega)
a_\lambda^\dag b(\omega)+{\cal W}_\lambda^\star (\omega) a_\lambda b^\dag(\omega)\right]+
\nonumber
\\
&+&\sum_\lambda [g_\lambda\, a_\lambda\sigma^++g_\lambda^\star\, a_\lambda^\dag \sigma^- ]\,.
\label{H_main}
\end{eqnarray}
Note, that the form of this Hamiltonian is a bit simpler as compared to the one
presented in \cite{VivPRA03} as we do not consider multiple scattering channels
outside the cavity. 

The Hermitian resonator modes are described by a discrete set of operators $a_{\lambda}$ and
corresponding eigenfrequencies $\omega_\lambda$, whereas the external radiation field 
corresponds to a continuous set of operators $b(\omega)$ and frequencies $\omega$. 
The operators obey the usual canonical
commutation relations (see Sec. IID in \cite{VivPRA03} for more details). The
resonator and external region communicate with each other via the coupling matrix
elements ${\cal W}_\lambda (\omega)$ defined as the expectation value of the operator
${\cal L}_{PQ}$ sandwiched between the resonator and external modes (see Eq.~(52a) in
\cite{VivPRA03}). This coupling operator is determined through the 
Feshbach projection formalism, which consists of separating space in two regions, the
resonator $Q$  and the external region $P$. Finally, the action of the operator
${\cal L}$ onto an arbitrary function $\phi$ is written as the decomposition ${\cal
L} \phi={\cal L}_{QQ} \mu+{\cal L}_{QP} \nu+{\cal L}_{PQ} \mu+{\cal L}_{PP} \nu$,
where the functions $\mu$ and $\nu$ reside inside the resonator and the external regions,
respectively. Correspondingly, the operators  ${\cal L}_{QP}$ and ${\cal L}_{PQ}$ act in the vicinity of the boundaries between the resonator and external region (see Sec.~IIB-C for more details). The key point is that the total operator ${\cal L}$ as well as the cavity
operator, ${\cal L}_{QQ}$, and external region operator, ${\cal L}_{PP}$, are {\it
Hermitian} operators in their regions of definition. The operators $\sigma_z$, $\sigma^+$ and $\sigma^-$ are
the standard Pauli operators which describe the TLS and $\omega_a$ stands for
its transition frequency. The coupling amplitude $g_\lambda$ is given by 
\begin{eqnarray}
\label{g_lambda}
g_\lambda=-i\left(\dfrac{\hbar\omega_\lambda}{2}\right)^{1/2} {\pmb \mu}
\cdot {\bf u}_\lambda ({\bf r_a}),
\end{eqnarray}
where ${\pmb \mu}$ is the dipole strength of the transition, ${\bf u}_\lambda ({\bf
r})$ stands for the eigenfunctions of ${\cal L}_{QQ}$ and ${\bf r_a}$ is the location
of the TLS.

It should be noted that in the Hamiltonian (\ref{H_main}) the rotating wave
approximation has already been applied in the following ways: (i) The nonresonant terms in the system-and-bath part of Hamiltonian [i.e., terms proportional to $a_\lambda^\dag
b^\dag(\omega)$ and $a_\lambda b(\omega)$] are neglected. This approximation is valid
if the damping rates of the cavity resonances are substantially smaller than the
frequencies of interest. For our purpose this approximation is, indeed, well fulfilled 
since the revival regime that we aim to describe occurs 
exactly in this limit; (ii) Also the nonresonant terms in the atom-field interaction [i.e.,
terms proportional to $a_\lambda^\dag\sigma^+$ and $a_\lambda \sigma^-$] are
neglected which is a commonly used approximation.

\subsection{Volterra equations}
Since the Hamiltonian, Eq.~(\ref{H_main}), conserves the total number of atom and field
excitations (thanks to the above rotating wave approximation) we can set up the following
ansatz for our solution to the Schr\"odinger equation
\begin{eqnarray}
\label{Ansatz_algemein}
&&|\Psi(t)\rangle=c(t)e^{-i \omega_a t/2} |u\rangle |0\rangle+
\\
\nonumber
&+&\sum_\lambda c_\lambda(t) |l\rangle |1_\lambda\rangle e^{-i(\omega_\lambda-\omega_a/2) t}+
\\
\nonumber
&+&
\int d\omega\, c(\omega,t) e^{-i(\omega-\omega_a/2) t} |l\rangle |1(\omega)\rangle,\,\,\,
\end{eqnarray}
where the ket-vectors $|u\rangle$ and $|l\rangle$ stand for the atom in the upper
and lower states respectively. In Eq.~(\ref{Ansatz_algemein}) the ket-vectors
$|0\rangle$, $ |1_\lambda\rangle$ and $|1(\omega)\rangle$ represent the vacuum state
of the electromagnetic field, a single photon in cavity mode $\lambda$ and a
single photon in the external region with frequency $\omega$, respectively.
We assume that the system at time $t=0$ is in the initial state $|u\rangle |0\rangle$. 
After straightforward algebra we derive the following set of coupled differential equations
for the probability amplitudes $c(t)$, $c_\lambda(t)$ and $c(\omega,t)$ introduced in
Eq.~(\ref{Ansatz_algemein}),
\begin{subequations}
\begin{eqnarray}
\label{Eq_c}
\dot c(t)&=&-\dfrac{i}{\hbar}\sum_\lambda g_\lambda
e^{-i(\omega_\lambda-\omega_a) t} \, c_\lambda(t) 
\\\nonumber
\\
\nonumber
\dot c_\lambda(t)&=&-\dfrac{i}{\hbar}g_\lambda^\star e^{i(\omega_\lambda-\omega_a) t}\, c(t) -
\\\nonumber
\\
&-&i\int d\omega {\cal W}_\lambda (\omega)e^{-i(\omega-\omega_\lambda) t}\, c(\omega,t)
\label{Eq_c_lambda}
\\\nonumber
\\
\dot c(\omega,t)&=&-i\sum_\lambda  {\cal W}_\lambda^\star (\omega)
e^{-i(\omega_\lambda-\omega) t} c_\lambda(t).
\label{Eq_c_omega}
\end{eqnarray}
\end{subequations}
The initial conditions are $c(0)=1$ and $c_\lambda(0)=c(\omega,0)=0$.

Next, we formally integrate Eq.~(\ref{Eq_c_omega}) and plug the result into 
Eq.~(\ref{Eq_c_lambda}) which allows us to exclude the external region from the
consideration, such that we finally obtain the following set of equations
\begin{subequations}
\begin{eqnarray}
\label{Eq_c_dot}
&&\dot c(t)=-\dfrac{i}{\hbar}\sum_\lambda g_\lambda 
e^{-i(\omega_\lambda-\omega_a) t} \, c_\lambda(t)
\\\nonumber
\\\nonumber
&&\dot c_\lambda(t)=-\dfrac{i}{\hbar}g_\lambda^\star e^{i(\omega_\lambda-\omega_a)
t} \, c(t)-\!\!
\int d\omega \sum_{\lambda'} {\cal W}_\lambda(\omega) {\cal
W}_{\lambda'}^\star(\omega)\times
\\
&&\times e^{-i(\omega-\omega_\lambda) t} \int\limits_0^t d\tau 
e^{-i(\omega_\lambda'-\omega) \tau}\,  c_{\lambda'}(\tau).\,\,\,\,\,\,\,
\label{Eq_c_lambda_dot}
\end{eqnarray}
\end{subequations}
\subsection{Markov approximation}

To simplify matters, we apply the so-called Markov approximation in Eq.~(\ref{Eq_c_lambda_dot})
with respect to the cavity amplitudes $c_\lambda(t)$ such that memory
effects with regard to the outcoupling to the external radiation field are disregarded. 
(Note that, most importantly, the memory effects within the cavity are still carried along.)
Specifically, we shift the initial time of integration to
$-\infty$, let $c_\lambda(t') \approx c_\lambda(t)$ and, assuming subsequent
integration with respect to $\omega$, make use of the following relation
\begin{eqnarray}
e^{-i(\omega-\omega_\lambda) t}\cdot\lim_{\sigma\rightarrow 0}\left.
\dfrac{e^{i(\omega-\omega_{\lambda'}-i\sigma)
\tau}}{\omega-\omega_{\lambda'}-i\sigma}\right\vert_{\tau=-\infty}^{\tau=t}
\rightarrow 
\\\nonumber
\rightarrow
e^{-i(\omega_{\lambda'}-\omega_\lambda) t}\left[{\cal
P}\left(\dfrac{1}{\omega-\omega_{\lambda'}}\right)
+i\pi\delta(\omega-\omega_{\lambda'})\right],
\end{eqnarray}
where ${\cal P}$ stands for the principal value. The differential equations for
$c(t)$ and $c_\lambda(t)$ are then as follows
\begin{subequations}
\begin{eqnarray}
\label{c_Markov}
\dot c(t)&=&-\dfrac{i}{\hbar}\sum_\lambda g_\lambda 
e^{-i(\omega_\lambda-\omega_a) t} \, c_\lambda(t)
\\
\label{c_lambda_Markov}
\dot c_\lambda(t)&=&-\dfrac{i}{\hbar}g_\lambda^\star e^{i(\omega_\lambda-\omega_a)
t} \, c(t)+
\\\nonumber
&+&\sum_{\lambda'} \Gamma_{\lambda \lambda'}(\omega_{\lambda'})
e^{-i(\omega_{\lambda'}-\omega_{\lambda})t}c_{\lambda'}(t),
\end{eqnarray}
\end{subequations}
where the matrix elements of the damping matrix $\Gamma_{\lambda \lambda'}$ are given by
\begin{eqnarray}
\label{Eq_Wlambda}
\Gamma_{\lambda \lambda'}(\omega_{\lambda'})&=&-\pi {\cal
W}_{\lambda}(\omega_{\lambda'}){\cal W}_{\lambda'}^\star(\omega_{\lambda'})+
\\\nonumber
&+&i {\cal P} \int d\omega
\dfrac{{\cal W}_{\lambda}(\omega){\cal W}_{\lambda'}^\star(\omega)}{\omega-\omega_{\lambda'}},
\end{eqnarray}
which should be calculated in a discrete set of eigenfrequencies $\omega_\lambda$
only. The second term in Eq.~(\ref{Eq_Wlambda}) is similar to a Lamb shift in that it accounts
for a shift of the cavity resonances in an open system with respect to the positions
in the corresponding closed system. Next, we formally integrate Eqs.~(\ref{c_Markov}, \ref{c_lambda_Markov}) and end up with a set of coupled integral Volterra equations
\begin{eqnarray}
\label{ct_eq}
\nonumber
&&c(t)\!=\!1\!-\!\dfrac{i}{\hbar^2}\sum_\lambda\dfrac{g_\lambda
g_\lambda^\star}{\omega_\lambda-\omega_a}\!\!\cdot\!\!\int\limits_0^t
d\tau\left[e^{-i(\omega_\lambda-\omega_a)(t-\tau)}\!-\!1\right]\cdot c(\tau)
\\\nonumber
\\\nonumber
&&+\dfrac{1}{\hbar} \sum_{\lambda \lambda'}\dfrac{g_\lambda \Gamma_{\lambda
\lambda'}(\omega_\lambda') }{\omega_\lambda-\omega_a}\int\limits_0^t
d\tau\left[e^{-i(\omega_\lambda-\omega_a)(t-\tau)}\!-\!1\right]\times
\\\nonumber
\\
&&\times e^{-i(\omega_\lambda'-\omega_a)\tau} c_\lambda'(\tau);
\\\nonumber
\\\nonumber
&&c_\lambda(t)=-\dfrac{i g_\lambda^\star}{\hbar} \int\limits_0^t d\tau
e^{i(\omega_\lambda-\omega_a) \tau} \, c(\tau)+
\\\nonumber
\\
&&+\sum_{\lambda'}\Gamma_{\lambda\lambda'}(\omega_\lambda') \int\limits_0^t d\tau
e^{-i(\omega_{\lambda'}-\omega_{\lambda})\tau}c_{\lambda'}(\tau).
\label{clambda_eq}
\end{eqnarray}
\begin{figure}
\includegraphics[angle=0,width=1.\columnwidth]{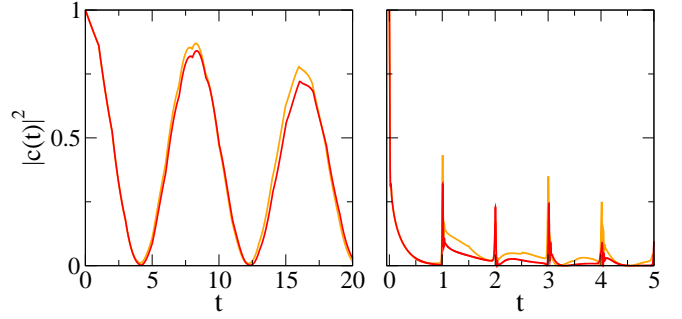}
\caption{Comparison between the results obtained from a single Volterra equation [red (dark gray) curves] and from the system-and-bath formalism [orange (gray) curves]. The calculations are performed for the 1D geometry presented in Fig.~\ref{fig_contour} with the mirror reflectivity parameter set to $\eta = 0.18$.  {\it Left panel}:  $\gamma=2.5 \cdot 10^{-3}$ (regime of Rabi oscillations). {\it Right panel}:  $\gamma=1.44$ (multimode strong coupling regime). Time $t$ is measured in units of
half the cavity round trip time.}
\label{fig7_Vives}
\end{figure}
\subsection{One-dimensional dielectric cavity} 

We solve Eqs.~(\ref{ct_eq}), (\ref{clambda_eq}) numerically for the geometry shown in Fig.~\ref{fig_contour}. Specifically, we consider the one-dimensional cavity of length $L$ now bounded at $x=-L,0$ by two thin semi-transparent mirrors modelled by dielectric slabs of width $d\ll L$ with refractivity index $n$. Using the fact that the TLS couples only to those modes which are symmetric with respect to the center of the cavity (where the TLS is located), we replace our original geometry by a more simple one. This new cavity runs from $[-L/2,0^-]$ with Neumann boundary conditions at the position of the TLS, $\partial_x u_{\lambda}(x=-L/2)=0$. On the right cavity edge we impose (for the closed system $Q$) a Dirichlet boundary condition, $u_\lambda(x=0^-)=0$, to remove a singular contribution of the operator ${\cal L}_{QQ}$ at this point (see, e.g., Eqs.~(52a) in \cite{Viviescas_geom})]. The corresponding cavity eigenvalue problem,
\begin{eqnarray}
\dfrac{d^2}{dx^2}u_{\lambda}(x)+ \omega_\lambda^2 u_{\lambda}(x)=0,
\end{eqnarray}
is finally solved with the eigenvalues $\omega_\lambda=\pi(2\lambda-1)/L$ ($\lambda=1,2,...$) and with the eigenvectors 
(inside the cavity)
\begin{eqnarray}
u_\lambda=\sqrt{\dfrac{2}{L}}\cos\left[\omega_\lambda \left(x+\dfrac{L}{2}\right) \right].
\end{eqnarray}
The coupling amplitudes between the TLS and the cavity modes (\ref{g_lambda}) reduce to
\begin{eqnarray}
\label{g_L_cav}
g_\lambda=i\mu \sqrt{\dfrac{\hbar\omega_\lambda}{L}}\cdot
f_c(\omega_\lambda).
\end{eqnarray}
In the limit of $n\rightarrow \infty$ and $d\rightarrow 0$, keeping the mirror's transparency factor $\eta=n^2 d$ finite, the channel modes 
(outside the resonator) coincide with those calculated in \cite{Viviescas_geom} [see Eqs.~(55-58) therein], 
\begin{eqnarray}
\nu(\omega,x)=\dfrac{1}{\sqrt{2\pi}}\left(e^{-i \omega x}+\dfrac{i-\eta \omega}{i+\eta \omega} e^{i \omega x}\right).
\end{eqnarray}
To couple these cavity modes in the bounded domain $Q$ to the unbounded domain $P$ we require the coupling elements
${\cal W}_{\lambda}(\omega)$ which enter the damping matrix $\Gamma_{\lambda \lambda'}$,
\begin{eqnarray}
{\cal W}_{\lambda}(\omega)=\dfrac{(-1)^{\lambda}}{1-i\eta \omega}\sqrt{\dfrac{\omega_\lambda}{\pi
\omega L}}\cdot f_c(\omega_\lambda).
\end{eqnarray}
Here and in Eq.~(\ref{g_L_cav}) we introduce the cut-off function $f_c(\omega_\lambda)= e^{-(\omega_{\lambda}-\omega_a)^2/(4 \omega_c^2)}$ to eliminate the interaction with high-frequency modes in the same way as was done in Sec.~\ref{Sec_Theor_Model}. To ensure the convergence of the integral in Eq.~(\ref{Eq_Wlambda}) also in the low frequency limit, we integrate from a frequency above zero but below the first cavity resonance. Finally, we plug the obtained expressions into  Eqs.~(\ref{Eq_Wlambda}-\ref{clambda_eq}) and solve them numerically with the initial conditions $c(0)=1$ and
$c_\lambda(0)=0$.  

The results of our calculations are shown in Fig.~\ref{fig7_Vives} for two typical values of the coupling strength within both the regime of Rabi oscillations and the regime of revivals. We normalize time to half the cavity round-trip time $L/c$ and find again the revivals occurring at integer multiples of these values. Note, in particular, the very good correspondence which we find between the results obtained from the model based on the CF state representation of the LDOPS within a single Volterra equation (\ref{Volterra_eq})  and the system-and-bath formalism given by Eqs.~(\ref{ct_eq}), (\ref{clambda_eq}) above. This close correspondence confirms the validity of our calculations and the difference in complexity between the two calculations demonstrates the usefulness of the simple and accessible approach presented in Sec.~\ref{Sec_Results}. 

\section{Conclusions and Outlook}
To summarize, we show how the emission process of a two-level atom changes as a function of its coupling strength to the electromagnetic field of an open multimode resonator. Solving the Volterra equation for the temporal decay through Laplace transform allows us to obtain the decay dynamics together with a corresponding graphical analysis which provides an intuitive understanding of the different regimes observed. On top of the familiar exponential decay and damped Rabi oscillations in the weak and strong coupling regime, respectively, we identify, for very strong coupling, a regime where the emitter couples to multiple modes, leading to pulsed revivals of its initial excitation. We expect that these predictions can be explicitly verified in various physical systems dealing with a two-level-like emitter inside an open multimode cavity. In particular, we have circuit QED setups in mind (e.g. \cite{Sapienza:2010vp,Srinivasan2007,Fink2008,Houck2008}), for which the coupling strength can be tuned by engineering the two-level system appropriately.

\section{Acknowledgements}

The authors would like to thank R. Luger, M. Malekakhlagh and C. Viviescas for helpful discussions. Financial support by the Vienna Science and Technology Fund (WWTF) through Project No. MA09-030 (LICOTOLI), the Austrian Science Fund (FWF) through Projects No. F25-P14 (SFB IR-ON) and No. F49-P10 (SFB NextLite), the National Science Foundation through the NSF CAREER Grant No. DMR-1151810 and the Swiss NSF through Grant No.PP00P2-123519/1 is gratefully acknowledged. We also profited from free access to the computational resources of the Vienna Scientific Cluster (VSC).

\appendix
\section{Laplace transform of the Volterra equation}
\label{App_Lapl}
\begin{figure}
\includegraphics[angle=0,width=.6\columnwidth]{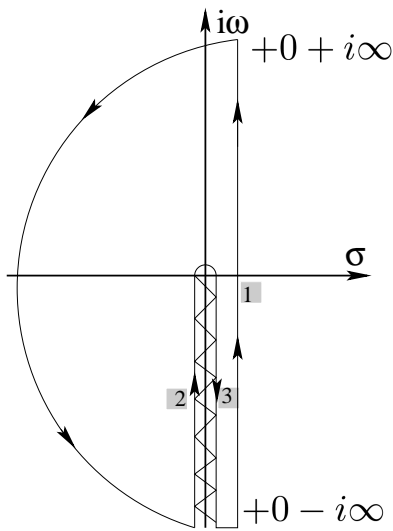}
\caption{Contour completion in the complex plane $s=\sigma+i \omega $ for the
calculation of the inverse Laplace transform, Eq.~(\ref{Inverse_LT}). Those
contours which give nonzero contribution are designated by numbers.}
\label{fig1_contour_supplem}
\vspace*{0.cm}
\end{figure}
We solve the Volterra equation (\ref{Volterra_eq}) by means of the standard Laplace transform method (see \cite{Riley} and also Chapter 5 in \cite{Berman10}, where different modal weight functions have been considered) multiplying it by $e^{-st}$ and integrating both sides of the equation with respect to time from $0$ to $\infty$. Here $s=\sigma+i\omega$ is the complex variable so that we reformulate our problem by solving it in the complex plane of $s$. After straightforward calculations, the algebraic equation for the Laplace transform, $\tilde c(s)=\int_0^\infty dt \,e^{-st} c(t)$, is derived which is solved by
\begin{equation}
\tilde c(s)=\dfrac{1}{s+\dfrac{\gamma}{\pi}\int_0^{\infty}d\omega
\dfrac{F(\omega)}{s+i(\omega-\omega_a)}}.
\end{equation}
Next, we perform the inverse Laplace transformation, $c(t)=\dfrac{1}{2\pi i}\int_{\sigma-i \infty}^{\sigma+i \infty}ds e^{st} \tilde c(s)$, and obtain the following formal solution for the amplitude $c(t)$
\begin{equation}
\label{Inverse_LT}
c(t)=\dfrac{e^{i\omega_a t}}{2\pi i}
\int_{\sigma-i \infty}^{\sigma+i \infty} \dfrac{e^{st}ds}{s+i\omega_a+G(s)},
\end{equation}
with
\begin{equation}
G(s)=\dfrac{\gamma}{\pi}\int_0^{\infty}\dfrac{d\omega F(\omega)}{s+i\omega}.
\end{equation}
where $\sigma>0$ should be chosen such that the real parts of all singularities of $\tilde c(s)$ are smaller than $\sigma$. It can be shown that the function 
\begin{equation}
J(\omega)=\lim_{\sigma\rightarrow 0^{+}}[G(\sigma+i \omega)-G(-\sigma+i \omega)]
\end{equation}
is nonzero for $-\infty<\omega \le 0$. Therefore the function $G(s)$, and as a consequence, the whole
integrand in Eq.~(\ref{Inverse_LT}) exhibits a jump along the negative part of the imaginary axis which is a branch cut. By equating the denominator of Eq.~(\ref{Inverse_LT}) to zero, $s+i\omega_a+G(s)=0$, the poles $s_j$ are shown to satisfy the following equation 
\begin{equation}
\label{Eq_poles}
\omega_j+\omega_a=\dfrac{\gamma}{\pi}\int_0^{\infty}d\omega
\dfrac{F(\omega)}{\omega+\omega_j},\,\,\,\sigma_j=0.
\end{equation}

Thus, the poles (if at all existing) can be located on the imaginary axis only. Moreover, we strictly prove using the graphical analysis and the fact that $F(\omega) \ge 0$, that only a single simple pole can reside in the positive imaginary axis which leads to undamped oscillations at infinite time. For values of the
coupling strength $\gamma$ larger than considered in this paper, such a scenario emerges in the equations but is not considered here. Thus, to evaluate the original integral, Eq.~(\ref{Inverse_LT}), we apply Cauchy's theorem to a closed contour shown in Fig.~\ref{fig1_contour_supplem}. We prove similarly to the Jordan's lemma, that the arc-contribution is negligible, and the contribution of the small semi-circle  around $s=0$ is also zero. Therefore, the only paths which remain are those around the branch cut 
and the one we are looking for, see Fig.~\ref{fig1_contour_supplem}. Thus, we derive the following expression for the amplitude $c(t)$
\begin{equation}
c(t)=\dfrac{e^{i\omega_a t}}{2\pi i} \int_{0}^{\infty} d\omega e^{-i\omega t} \left(\Phi_{-}(\omega)-\Phi_{+}(\omega)\right),
\end{equation}
where
\begin{eqnarray}
\label{Eq_Phis_34}
&&\Phi_{\pm}(\omega)=
\\
\nonumber
&&\lim_{\sigma\rightarrow 0^{+}}\left\{\dfrac{1}
{\omega-\omega_a+i\left[\dfrac{\gamma}{\pi} \int_0^{\infty}\dfrac{d\tilde\omega F(\tilde\omega)}{\pm \sigma+i(\tilde\omega-\omega)}
\pm \sigma\right]}\right\}.
\end{eqnarray}
Employing the Sokhotski-Plemelj theorem, the integral in the denominator of Eq.~(\ref{Eq_Phis_34}) is rewritten in the limit of $\sigma\rightarrow 0$ as
\begin{equation}
\nonumber
\int_0^{\infty}\dfrac{d\tilde\omega F(\tilde\omega)}{\pm
\sigma+i(\tilde\omega-\omega)}=
-i\left\{\mathcal{P}\int_0^{\infty}\dfrac{d\tilde\omega F(\tilde\omega)}{\tilde\omega-
\omega}\pm i\pi F(\omega)\right\},
\end{equation}
where $\mathcal{P}$ denotes the Cauchy principal value. We finally end up with Eqs.~(\ref{Eq_fin_c}-\ref{Lamb_shift}) for the amplitude $c(t)$ (see Sec.~\ref{Sec_Results}).

\end{document}